\documentclass[12pt]{article}

\newcommand{\be}{\begin{equation}}
\newcommand{\ee}{\end{equation}}
\newcommand{\bea}{\begin{eqnarray}}
\newcommand{\eea}{\end{eqnarray}}

\begin{document}
\global\parskip 6pt

\begin{titlepage}

\begin{center}
{\Large\bf Self-T-dual Brane Cosmology}\\

\vspace*{1cm} Massimiliano Rinaldi\\

\vspace*{1cm} {School of Mathematical Sciences, University College Dublin, \\
Belfield, Dublin 4, Ireland, \\
and\\
Dipartimento di Fisica and I.N.F.N, Universit\`{a} di Bologna, \\
Via Irnerio 46, 40126 Bologna, Italy.\\\vspace*{0.5cm}
E-mail: {\verb"rinaldim@bo.infn.it"}}\\

\vspace*{1cm}

\begin{abstract}
\noindent We show how T-duality can be implemented with brane
cosmology. As a result, we obtain a smooth bouncing cosmology with
features similar to the ones of the pre-Big Bang scenario. Also, by
allowing T-duality transformations along the time-like direction, we
find a static solution that displays an interesting self tuning
property. \vspace*{2.5cm}

\noindent {\it Based on a talk given at the XI Marcel Grossmann
Meeting, Berlin 23-29 July, 2006.}
\end{abstract}
\end{center}
\end{titlepage}

\section{Introduction}

In the past years, various cosmological models were inspired by
different aspects of string theory. In some cases, these rely upon
the fundamental symmetries of string theory, the most notable
example being the pre-Big Bang (PBB) scenario motivated by
T-duality~\cite{Gasperini}. In other cases, models are based on
extended objects such as branes~\cite{Horava}. These two approaches
are often seen as competing. However, if our Universe is seen as a
brane moving in a higher-dimensional bulk, obtained by
compactification of string theory, it is likely that the effective
cosmology inherits some of the symmetries of the uncompactified
theory. A first application of this idea can be found in the context
of type IIA/IIB supergravity. When compactified to five dimensions,
these theories possess static black hole solutions with flat
horizon~\cite{maxBH}, which are directly related by T-duality
transformations. By studying a brane moving in these dual spaces, it
was found that these transformations induce the inversion of the
cosmological scale factor on the brane, along the lines of the PBB
scenario~\cite{max}. The latter, however, is based on a self-T-dual
action, with time-dependent background solutions. In the next
section, we show that it is possible to construct a self-T-dual
action, which, instead, has static background solutions. Also, an
embedded moving brane displays an effective cosmological evolution,
which smoothly connects a pre- and a post-big bang phase, through a
non-singular bounce, in complete analogy with some of the PBB
models. Finally, in the last section, we will also show how
Self-T-dual brane models address the problem of fine-tuning between
the brane vacuum energy density (tension) and the bulk cosmological
constant.

\section{Pre-Big Bang on the brane}

To recreate a PBB scenario on the brane, we must find first a
self-T-dual action, such that the related equations of motion have
static solutions. Let us consider the dilaton-gravity
action~\cite{maxpaul} \bea\label{action} S_{\rm bulk}=\int_{\cal
M}d^5x\sqrt{g}\,e^{-2\phi}\left[{\cal R}+4(\nabla \phi)^2+V\right]~,
\eea where $V$ is an exponential function of the so-called shifted
dilaton $\bar\phi$. If we choose the line element \bea\label{metric}
ds^2=-A^2(r)\,dt^2+B^2(r)\,dr^2+R^2(r)\, \delta_{ij}\,dx^i\,dx^j~,
\eea then the shifted dilaton is defined as
$\bar\phi(r)=\phi(r)-\frac{3}{2}\ln R(r)$. It can be shown that the
action (\ref{action}) is invariant under the T-duality
transformation $R(r)\stackrel{\rm T}{\longrightarrow}R(r)^{-1}$,
which leaves the shifted dilaton unchanged. Therefore, to any
solution with metric (\ref{metric}), there exists another with $R$
replaced by $1/R$. This property holds if we neglect the boundary
terms springing from variation of the action with respect to the
fields. However, if we want to preserve self-T-duality, these terms
must be kept when we introduce a $Z_2$ symmetric 3-brane, which acts
as a boundary. In this way, it turns out that the full action,
obtained by adding Eq.~(\ref{action}) to the brane action
\begin{equation}\label{extraterm} S_{\rm
brane}=-\int_{\Sigma}d^3x\,d\tau\sqrt{h}e^{-2\phi}\left[4K+{\cal
L}\right]~,
\end{equation} is still invariant under the transformation
$R(r)\stackrel{\rm T}{\longrightarrow}R(r)^{-1}$, {\it provided}
${\cal L}\rightarrow {\cal L}$, i.e. provided the brane matter
Lagrangian is itself T-duality invariant. In the expression above,
$h$ is the determinant of the induced FLRW metric on the brane,
$~ds^2=-d\tau^2+R^2(r(\tau))\delta_{ij}\,dx^i\,dx^j~$, $K$ is the
trace of the brane extrinsic curvature, and $\tau$ is the proper
cosmological time (and the parametric the position of the brane in
the bulk). It is clear that the duality acting on the bulk metric
leads to the inversion of the scale factor $R$. Now, let the matter
on the brane be a perfect fluid, with equation of state
$p=\omega\rho$. By carefully studying the Israel junction
conditions, it can be shown that the self-T-duality of ${\cal L}$
implies that $\omega\stackrel{\rm T}{\longrightarrow} -\omega$,
exactly like in the PBB scenario~\footnote{It is important to remark
that in this model, $\omega$ is an function of $\tau$.}.

By studying the bulk equations of motion, one can find black hole
solutions with one regular horizon. A brane moving in such a
background encounters a turning point outside the horizon. By
assuming that the T-duality transition occurs at the bounce, one can
construct a non-singular transition between a pre-big bang phase
(with, say, $-\omega$ and scale factor $1/R(\tau)$) and post-big
bang phase (with $\omega$ and scale factor $R(\tau)$). Finally, it
also turns out that the cosmic evolution far away from the bounce,
both in the past and in the future, is always of de Sitter type.

\section{Time-like T-duality and the fine tuning problem}

We now turn our attention to the fine tuning problem of
Randall-Sundrum-like models. We consider again the action
(\ref{action}), but now we assume that the bulk metric has the
Poincar\'{e}-invariant form\bea
ds^2=e^{2\sigma(z)}\eta_{\mu\nu}dx^{\mu}dx^{\nu}+dz^2~,\eea while
the shifted dilaton reads $\bar\phi=\phi-2\sigma$. Along the lines
sketched in the previous section, one can show~\cite{maxollo} that
the action is invariant under T-duality transformations along both
the time and space coordinates $x^{\mu}$, i.e. under
$\sigma(z)\stackrel{\rm T}{\longrightarrow} -\sigma(z)$. Of course,
the same holds for the equations of motion, for which the only
non-singular solutions are given by a constant shifted dilaton
$\bar\phi_0$ and $\sigma=\pm\lambda(z-z_0)$, where $\lambda$ and
$z_0$ are integration constants. The positive and negative sign
solutions are related by T-duality, and they simply correspond to
two slices of anti-de Sitter space~\footnote{This is consistent with
time-like T-duality, which requires the time-like direction to be
compact.}. By inserting a $Z_2$-symmetric brane on this background,
with a perfect fluid as matter, we can preserve the self-T-duality
of the total action provided we impose $(\omega+1)\stackrel{\rm
T}{\longrightarrow}-(\omega+1)$. Interestingly, this duality
transformation is identical to the one found in the context of
phantom cosmology~\footnote{Thanks to Prof. M.~P.~Dabrowski for
pointing out this similarity~\cite{dab}.}. But the most important
result is that, in the case of a static brane, the (constant) energy
density of the brane matter is given by \bea \rho^2=
4V_0\,e^{\,\beta\bar\phi_0}~, \eea where $\beta$ and $V_0$ are
arbitrary constants. Therefore, any value of $\rho$ can be reached
given any vacuum expectation value of the shifted dilaton.

These encouraging results call for further investigations into
self-T-dual brane cosmology, and the main target is to find some
signatures (such as particular CMB fluctuations or relic gravitons)
of this model, which might be tested by observations.

\section*{Acknowledgments} I wish to thank P.~Watts and O.~Corradini
for their fundamental contributions to these results, and
Prof.~D.~Gal'tsov for inviting me to speak at the parallel session.


\end{document}